\documentclass[a4paper]{nature}

\usepackage{bm}
\usepackage{authblk}
\usepackage{amsmath}
\usepackage{graphicx}
\usepackage{xcolor}
\usepackage[section]{placeins}
\makeatletter
\let\saved@includegraphics\includegraphics
\AtBeginDocument{\let\includegraphics\saved@includegraphics}
\renewenvironment*{figure}{\@float{figure}}{\end@float}
\makeatother
\usepackage[labelfont=bf]{caption}
\usepackage{cite}
\usepackage{soul}
\usepackage[normalem]{ulem}
\DeclareUnicodeCharacter{FF0C}{,}
\newcommand{\upcite}[1]{\textsuperscript{\scalebox{0.7}{\cite{#1}}}}

\title{Room-temperature ferroelectric $p$-wave magnet \footnotetext{$^\ast$  E-mail: chenxianzhe@fudan.edu.cn, lsmejkal@pks.mpg.de
\\}}

\author[1$\dagger$]{Kaiwen Shen}
\author[2$\dagger$]{Nicoly Hembeck}
\author[2$\dagger$]{Warlley H. Campos}
\author[4$\dagger$]{Jianting Dong}
\author[5$\dagger$]{Bin Li}
\author[6]{Yuhang Li}
\author[1]{Gui Wang}
\author[2]{Jan Priessnitz}
\author[1]{Yifan Gao}
\author[1]{Yingfen Wei}
\author[1]{Hao Jiang}
\author[1]{Xumeng Zhang}
\author[1]{Ming Wang}
\author[2]{Jia Zhang}
\author[5]{Wu Shi}
\author[1]{Qi Liu}
\author[7,8$\ast$]{Libor \v{S}mejkal}
\author[1$\ast$]{Xianzhe Chen}
\author[1]{Ming Liu}

\affil[1]{Frontier Institute of Chip and System, State Key Laboratory of Integrated Chips and Systems, Zhangjiang Fudan International Innovation Center, Fudan University, Shanghai, 200433, China}
\affil[2]{Max Planck Institute for the Physics of Complex Systems, N\"othnitzer Str. 38, 01187 Dresden, Germany}
\affil[3]{School of Natural Sciences, Technische Universität München, 85748 Garching, Germany}
\affil[4]{School of Physics and Wuhan National High Magnetic Field Center, Huazhong University of Science and Technology, 430074 Wuhan, China}
\affil[5]{State Key Laboratory of Surface Physics and Institute for Nanoelectronic Devic    es and Quantum Computing, Fudan University, Shanghai 200433, China}
\affil[6]{School of Physics, Nankai University, Tianjin 300071, China}
\affil[7]{Max Planck Institute for Chemical Physics of Solids, N\"othnitzer Str. 40, 01187 Dresden, Germany}
\affil[8]{Institute of Physics, Czech Academy of Sciences, Cukrovarnick\'a 10, 162 00 Praha 6, Czech Republic}
\graphicspath{{figures/}}

\date{}                   
\begin{document}
	
	\maketitle
	
	\vspace{10pt}
	
	\begin{abstract}
		\textbf{
$P$-wave magnets exhibit odd-parity spin polarization in momentum space while retaining zero net magnetization, offering a promising route toward nonrelativistic spintronics\upcite{jungwirth2026,hellenes2024,brekke2024,ezawa2024,ezawa2025,hodt2025,maeda2024,song2025,yamada2025,Priessnitz2026}. Previously reported $p$-wave magnets based on combined time-reversal and lattice-translation symmetry, $\mathcal{T}\mathbf{t}$, have thus far exhibited ordering temperatures below the liquid-nitrogen range \upcite{song2025}. Here, we demonstrate a distinct route to a pristine $p$-wave magnetism that does not require $\mathcal{T}\mathbf{t}$ symmetry, and instead is based on incommensurate spin-spiral order. Specifically, the cycloidal magnetic structure of the multiferroic BiFeO$_3$ (BFO) intrinsically realizes $p$-wave magnetism coexisting with ferroelectric polarization at room temperature. Combining symmetry analysis, density-functional-theory calculations, and circular photogalvanic-effect (CPGE) measurements, we find a large nonrelativistic spin splitting of approximately $100$~meV associated with the cycloidal order. The resulting $p$-wave spin polarization gives rise to a characteristic CPGE response whose sign can be reversibly switched by electrically controlling the cycloid chirality. Our results establish ferroelectric BiFeO$_3$ as a room-temperature platform for electrically switchable $p$-wave magnetism and provide a route toward voltage-controlled spintronic devices based on parity-breaking magnetic order \upcite{Priessnitz2026}.}\\
	\end{abstract}

    The discovery of magnetic states with vanishing net magnetization but robust non-relativistic spin splitting has revolutionized our understanding of magnetic symmetry and its potential for spintronics\upcite{ahn2019,feng2022,smejkal2023,reichlova2024,amin2024,krempasky2024,zhang2024,bhowal2024,chen2025,duan2025,gu2025}. Beyond the recently categorized altermagnets with  even-parity ($d$, $g$, $i$-wave) collinear spin splitting, odd-parity magnets with collinear $p, f, ...$-wave spin splitting---also called antialtermagnets~\upcite{jungwirth2025altermagnetism}---
    offer a solid-state analog to the parity-breaking excitations in ${}^3$He superfluids\upcite{jungwirth2025}. %
    % {To date, $p$-wave magnets  have been realized in materials with spin helix at cryogenic temperature, such as NiI${}_2$ and Gd${}_3$Ru${}_4$Al${}_{12}$. %, which are strictly limited by the even‑periodicity condition.}
    %$P$-wave magnets with ferromagnetic helix  have only been realized in cryogenic materials such as NiI${}_2$, Gd${}_3$Ru${}_4$Al${}_{12}$, and are strictly limited by the even‑periodicity condition, leaving almost no scalable room‑temperature candidates. 
    To date, signatures of $p$-wave magnets have been reported in materials with spin helix at cryogenic temperature. Song {\textit{et al.}} reported the electrical switching of the spin polarization consistent with $p$-wave magnetism in NiI${}_2$ via the circular photogalvanic effect (CPGE)\upcite{song2025}, while Yamada {\textit{et al.}} reported magnetic state in a metallic helimagnet Gd${}_3$Ru${}_4$Al${}_{12}$\upcite{yamada2025} supporting $p$-wave spin splitting and under weak modulation of the magnetic order a giant anomalous Hall effect. 
     In $p$-wave magnets with spin cycloid or helix, the space-time inversion $\mathcal{PT}$ and  $\mathcal{P}$ symmetries are broken. 
     This, together with the $\mathcal{T}\mathbf{t}$ symmetry present  in archetypal commensurate $p$-wave magnets, ensures odd-parity time-reversal-symmetric spin polarization. In addition, the coplanar spin-only group $Z_2^{C_{2\perp}\mathcal{T}}$\upcite{elcoro2026spin} enforces the collinearity of the spin-polarization~\upcite{hellenes2024}. %
     % 
     %In addition, the coplanar spin-order is invariant under the spin-space symmetry $C_{2\perp}\mathcal{T}$~\upcite{elcoro2026spin}, where $C_{2\perp}$ is a $180^\circ$ spin-only rotation perpendicular to the spiral plane, which enforces the collinearity of the spin-polarization~\upcite{hellenes2024}.

     The helicoidal spiral-like spin order was broadly studied in BiFeO${}_3$ (BFO). BFO is a quintessential type-I multiferroic with a high N\'eel temperature ($T_N\sim640~\mathrm{K}$) and robust ferroelectricity ($T_C=1100~\mathrm{K}$, polarization  $90~\mu\mathrm{C}\cdot\mathrm{cm}^{-2}$). BFO has been intensively studied for more than two decades as a canonical model system for understanding strong magnetoelectric coupling and electric field control of magnetism\upcite{bhattacharjee2014ultrafast,meisenheimer2024switching,pal2026bulk,wang2026,husain2026}. Below $T_N$, BFO realizes an incommensurate magnetic cycloidal texture incompatible with $\mathcal{T}\mathbf{t}$ symmetry. The reversal of ferroelectric polarization and DMI direction can reverse the chirality of spin cycloids from left-handness to right-handness\upcite{huang2024}, rendering BFO a leading candidate for low power memory and logic applications\upcite{ederer2005,albrecht2010,rahmedov2012,manipatruni2019,caretta2023,li2023,meisenheimer2024,chai2024,huang2024}. %Despite extensive research over the past two decades, the intrinsic $p$-wave magnetic nature of this cycloidal order has remained elusive. 

\begin{figure}[t!]
    	\centering
    	\includegraphics[width=1\textwidth]{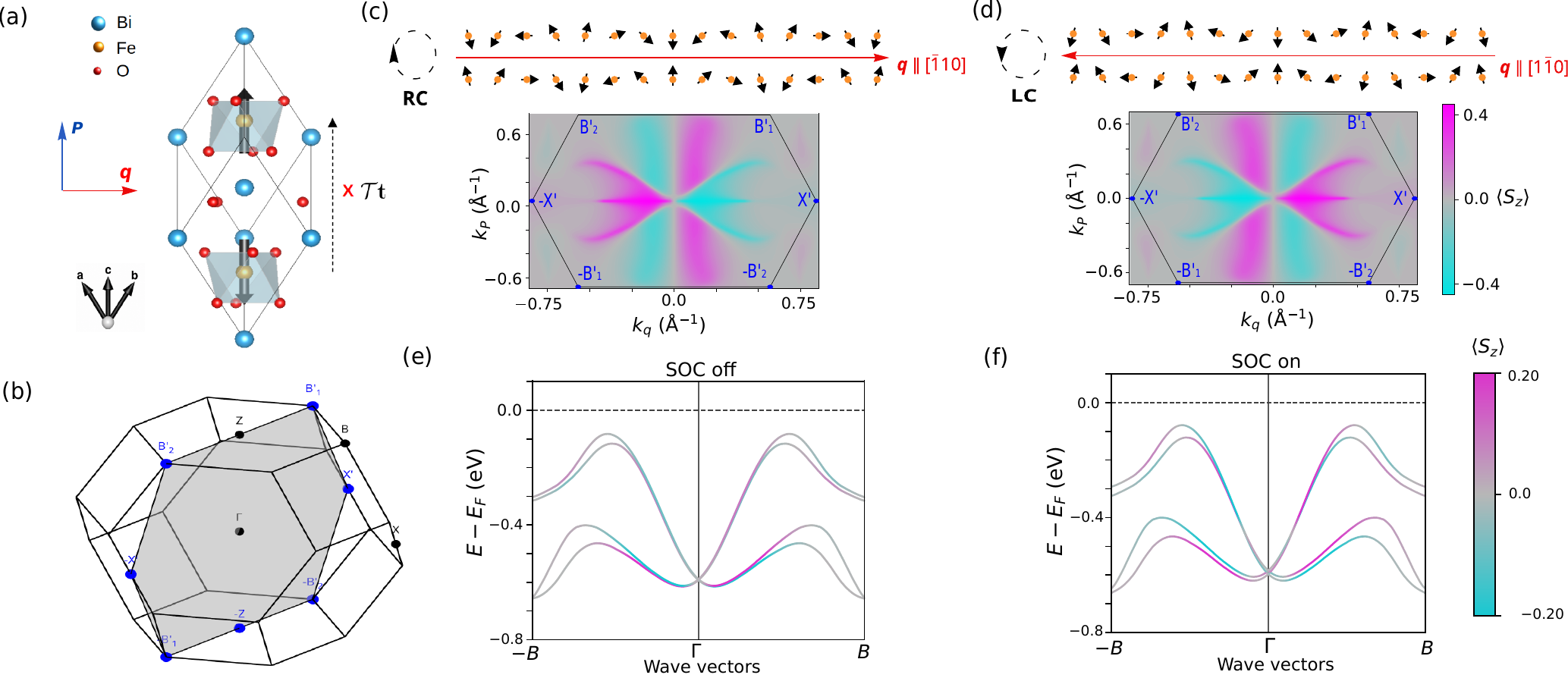}
    	\caption{\textbf{Ferroelectric $p$-wave magnetism in incommensurate BiFeO${}_3$ spin cycloids. a, }Schematics of the BiFeO$_3$ rhombohedral unit cell, indicating the ferroelectric polarization $\mathbf{P}$, the cycloidal propagation vector $\mathbf{q}$, and the broken $\mathcal{T}\mathbf{t}$ symmetry. 
        \textbf{b,} First Brillouin zone of BiFeO${}_3$ parent crystal structure. The marked gray plane is parallel to $\mathbf{P}$, $\mathbf{q}$, and to the plane of the cycloid.
        \textbf{c,d,} Illustration of the incommensurate spin cycloids in BiFeO${}_3$ for right-handed chirality (RC) and left-handed chirality (LC) with the corresponding $p$-wave spin polarizations calculated via density functional theory. \textbf{e,} Odd-parity spin-polarized electronic band structure of BiFeO${}_3$ for RC cycloid computed without spin-orbit coupling (SOC). \textbf{f,} The same as \textbf{e}, but computed with SOC.
        } 
    	\label{fig:1}
    	%\vspace{80pt}
    \end{figure}

    Here, we demonstrate that this incommensurate cycloidal spin texture provides a previously overlooked mechanism for generating $p$-wave spin polarization at room temperature, which additionally  coexists and can interplay with the ferroelectricity of BFO. %(Extended Data Figure 1). %Distinct from both Rashba–SOC systems and p‑wave magnets with ferromagnetic spin helix, this antiferromagnetic $p$‑wave state exhibits giant non‑relativistic spin splitting (~130 meV) driven purely by chiral antiferromagnetic exchange interactions, not spin–orbit coupling. This represents a new mechanism for odd‑parity spin‑momentum locking in solids, unique to antiferromagnetic $p$‑wave order.
    %{\color{blue}
    % Unlike conventional Rashba systems where spin-orbit coupling (SOC) is mandatory, the $p$-wave spin-splitting in BFO originates from the non-relativistic exchange interaction within its chiral antiferromagnetic lattice of cycloid. 
    % The emergence of $p$-wave magnetism in BFO is rooted in its unique magnetic symmetry.
    As illustrated in Fig.~\ref{fig:1}a, BFO crystalizes in a rhombohedral structure with lattice constant $a_0 = 5.63195$~\AA~and ferroelectric polarization $\mathbf{P}\parallel[111]$ compatible with the broken $\mathcal{P}$ symmetry of its parent space group R3c (\# 161).  Fig.~\ref{fig:1}b shows the first Brillouin zone (BZ) of the rhombohedral unit cell and a grey-shaded plane marks the plane of the spin cycloid.  The two magnetically compensated Fe sublattices break $\mathcal{T}\mathbf{t}$ symmetry and form the reference magnetic motif of the cycloid, which is propagated through the crystal by $\mathbf{q}$-dependent rotations of the magnetic moments, with $\mathbf{q}=(-0.0045,0.0045,0)$ in reciprocal-lattice units. %As shown in Fig.~\ref{fig:1}b, the cycloidal order, characterized by a propagation vector $\mathbf{q}$ along the [$\overline{1}10$] direction, features a $\mathcal{T}\tau$ symmetry. 
    %Here, $L$ is the period of the cycloid in unit of the lattice constant $a_0$, giving $La_0\approx 62$~nm. Crucially, since the two magnetic sublattices are reversed, the condition $\mathcal{T}\tau$ symmetry holds irrespective of whether $L$ is odd or even.

    %This differs from the case of $p$-wave magnet with ferromagnetic spin helix, where $\mathcal{T}\tau$ symmetry is maintained only for even $L$.

    To demonstrate $p$-wave spin polarization in BFO, we performed density functional theory (DFT) calculations within the generalized Bloch theorem for magnetic spirals (see Methods). 
    %(SI: In our calculations, the slight canting between the magnetic moments of the two sublattices can be neglected, as the period of cycloid is about two orders of magnitude larger than the inter-sublattice distance.) 
     Fig.~\ref{fig:1}c shows the nonrelativistic spin expectation value $\langle S_z\rangle$ of the two spin-split top-most valence bands throughout the grey-shaded BZ cut for the cycloid with right-handed chirality (RC) in the top panel, corresponding to $\mathbf{q}||[\bar{1}10]$. In our spin coordinate system, the $S_z$ component is aligned perpendicular to the spin spiral. For clarity, from now on we will use the dimensionless out-of-plane spin operator $\sigma_\perp$, defined through the relation $S_z = \frac{\hbar}{2} \sigma_\perp$. Our DFT calculations reveal a collinear, out-of-plane spin polarization with a clear $p$-wave form.

The symmetry origin of this unconventional spin texture can be understood from the coplanar character of the incommensurate helicoidal order. In the absence of SOC, the magnetic structures exhibits the spin-only symmetry $C_{2\perp}\mathcal{T}$, where $C_{2\perp}$ denotes a
twofold rotation about the axis perpendicular to the spin plane. This
symmetry constrains the out-of-plane spin component to be odd in momentum~\upcite{hellenes2024,elcoro2026spin}, $C_{2\perp}\mathcal{T}\sigma_\perp(\boldsymbol{k})=-\sigma_\perp(-\boldsymbol{k})$.

%\textcolor{red}{\begin{equation}
%C_{2\perp}\mathcal{T}
%(\sigma_\parallel,\sigma_\perp)(\boldsymbol{k})
%=
%(\sigma_\parallel,-\sigma_\perp)(-\boldsymbol{k}),
%\end{equation}}%
%%
%where $\sigma_\parallel$ is the in-plane component of the dimensionless spin operator.

% \textcolor{red}{\begin{equation}
% C_{2\perp}\mathcal{T}
% (S_x,S_y,S_z)(\boldsymbol{k})
% =
% (S_x,S_y,-S_z)(-\boldsymbol{k}).
% \end{equation}}

% We now discuss the origin of this unconventional out-of-plane and
% $p$-wave spin polarization \textcolor{blue}{observed in our DFT calculations of BFO.
% Consistent with the calculations presented in Fig.~1c, we find a
% pronounced spin polarization in momentum space that is oriented
% perpendicular to the coplanar spin order in real space [***REDUNDANT WITH PREVIOUS PARAGRAPH***]}. This behavior
% can be understood from symmetry considerations. The incommensurate
% helicoidal order is coplanar and, in the absence of SOC, exhibits the
% spin-only symmetry $C_{2\perp}\mathcal{T}$---where $C_{2\perp}$ denotes a
% twofold rotation about an axis perpendicular to the spin plane. This
% symmetry constrains the out-of-plane spin component to have odd parity,
% \begin{equation}
% C_{2\perp}\mathcal{T}
% (S_x,S_y,S_z)(\boldsymbol{k})
% =
% (S_x,S_y,-S_z)(-\boldsymbol{k}).
% \end{equation}

Microscopically, for both commensurate and incommensurate spin
structures, this behavior can be understood from the relation
\cite{mitscherling2026microscopic}
\begin{equation}
\sigma_\perp \sim \left(\mathbf{m}_i \times \mathbf{m}_j\right)_\perp,
\end{equation}
where $\mathbf{m}_i$ and $\mathbf{m}_j$ denote the local
magnetic moments on sites $i$ and $j$, respectively. This relation
shows that electron hopping between sites $i$ and $j$ generates an
out-of-plane spin component proportional to the noncollinearity of the
corresponding local moments. Consequently, reversing the spin chirality from right to left-handed reverses the sign of
the resulting $p$-wave spin polarization. This chirality-dependent reversal is
confirmed by our DFT calculations in Figs.~\ref{fig:1}c,d for the right-handed chirality (RC) and left-handed chirality (LC), corresponding to $\textbf{q}||[\bar{1}10]$ and $\textbf{q}||[1\bar{1}0]$, respectively.

%Furthermore, the in-plane spin components vanish, despite the breaking of time-reversal symmetry $\mathcal{T}$. This can be understood as a consequence of the incommensurate nature of the spin order. For every local spin orientation along the propagation direction, there exists an oppositely oriented spin configuration elsewhere in the incommensurate texture. Upon integration over the spin texture, the contributions to both in-plane spin components therefore cancel.

Finally, we discuss the electronic band structure and the effect of
SOC. Figure~1e shows the insulating band structure and the two highest spin-split
valence bands, which exhibit large nonrelativistic spin splittings of
approximately $100$~meV. Upon inclusion of SOC, only minor
modifications of both the spin-splitting dispersion and the magnitude
of the spin polarization are observed, as shown in Fig.~1f. These
results corroborate the predominantly (nonrelativistic) exchange driven origin of the
$p$-wave magnetism.

\begin{figure}[t!]
    	\centering
    	\includegraphics[width=1\textwidth]{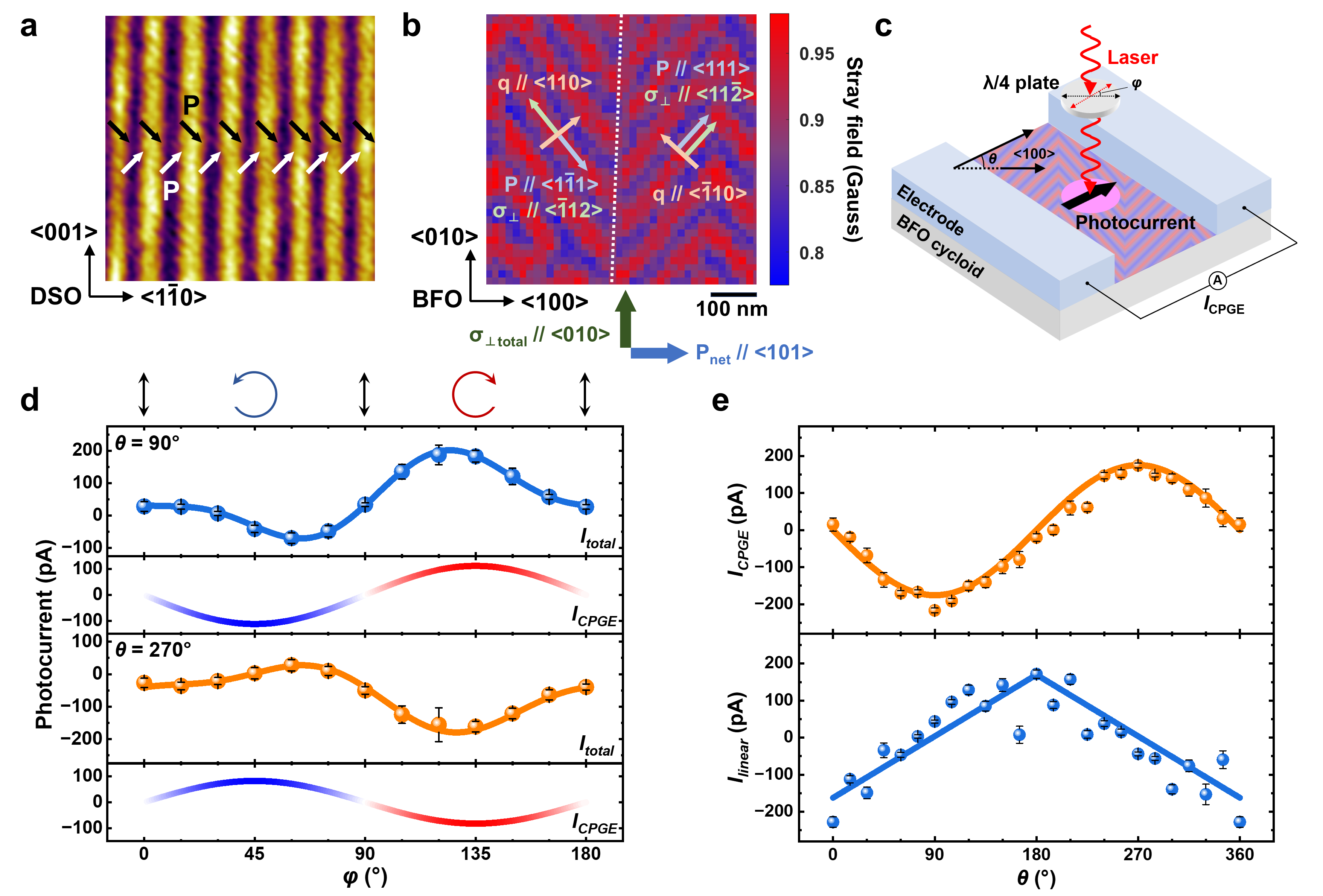}
    	\caption{\textbf{Magnetic chiral structure of BiFeO${}_3$ and corresponding circular photogalvanic effect (CPGE) signatures. a, }Lateral piezoresponse force microscopy (LPFM) image of the BFO film grown on a [110] oriented DSO substrate. The black and white arrows indicate the in-plane projections of the ferroelectric polarization vectors in two different directions in adjacent ferroelectric domains. \textbf{b,}  Magnetic domain image via the scan nitrogen vacancy microscope (SNVM) of the spin cycloid in BFO. Blue, red and green arrows represent the polarization direction $\mathbf{P}$, the propagating direction $\mathbf{q}$ and the spin polarization perpendicular to the spin-spiral plane $\langle\sigma_{\perp}\rangle$, respectively, within a single domain. White dashed lines indicate the domain boundary. The magnetic chirality is clockwise in the left domain and counter-clockwise in the right domain. The total $P_{net}$ and $\sigma_{\perp{total}}$ are represented by wider green and blue arrows, respectively, below the image. \textbf{c, }Experimental setup for CPGE effect detection. A 405~nm laser polarized by a $\lambda/4$ plate irradiates the surface of BFO, inducing a photocurrent which is captured by parallel detector electrode strips. $\varphi$ denotes the angle between the laser’s initial polarization direction and the optical axis of the plate. $\theta$ represents the angle between the photocurrent orientation and the [100] crystal direction of BFO. \textbf{d, }Photocurrent versus $\varphi$ for opposite $\theta$ with the maximum spin photocurrent $I_{CPGE}$. \textbf{e, } Spin photocurrent $I_{CPGE}$ and linear polarization contribution $I_{linear}$ variations with $\theta$. Peak $I_{CPGE}$ appears at $\theta{=90^{\circ}}$ and $\theta{=270^{\circ}}$, while peak $I_{linear}$ appears at $\theta{=0^{\circ}}$ and $\theta{=180^{\circ}}$. 
        } 
    	\label{fig:2}
    	%\vspace{80pt}
    \end{figure}

    To experimentally confirm $p$-wave magnetism from the cycloidal spin order in BFO and to clarify the direction of its spin splitting, we perform a combination of Piezoresponse Force Microscopy (PFM), NV center magnetic imaging, and anisotropic circular photogalvanic effect (CPGE) measurements, as shown in Fig.~\ref{fig:2}. Fig.~\ref{fig:2}a shows PFM domain structure of BFO, clearly resolving the $71^\circ$ ferroelectric domain structure—one of the most typical domain configurations in rhombohedral BFO. The bright and dark contrasts correspond to ferroelectric domains with distinct ferroelectric polarization directions ($\mathbf{P}$), as denoted by the small arrows. Specifically, the in-plane projection of the polarization direction of the bright domains is oriented $45^\circ$ relative to the [$1\overline{1}0$] direction of DSO substrate, while the dark domains exhibit a polarization direction rotated by $71^\circ$ relative to the bright domains. 
    
    The NV center magnetic imaging results in Fig.~\ref{fig:2}b demonstrates stray field mapping of a part of BFO region in Fig.~\ref{fig:1}a, directly visualizing the magnetic moment distribution and rotation behavior within individual $71^\circ$ ferroelectric domains. The color contrast in the image (from blue to red) represents the magnitude of the equivalent stray field induced by the cycloidal spin order, while the inset arrows indicate the direction of the equivalent $p$-wave spin polarization in each domain. For the dark ferroelectric domain in Fig.~\ref{fig:2}a, the ferroelectric polarization $\mathbf{P}$ is along the [$111$] direction, the magnetic domain precession direction $\mathbf{q}$ is along the [$\overline{1}10$] direction, and based on the symmetry analysis of Fig.~\ref{fig:1}a, the equivalent spin splitting direction is along the [$11\overline{2}$] direction. In contrast, for the adjacent bright $71^\circ$ domain, the ferroelectric polarization $\mathbf{P}$, the magnetic domain precession direction $\mathbf{q}$, and the equivalent spin polarization direction $\langle\sigma_{\perp}\rangle$ are reversed to the [$1\overline{1}1$]， [$110$]， [$\overline{1}12$] directions, respectively. Therefore, since the net ferroelectric polarization is aligned along [$101$] directions, the effective net spin splitting $\langle\sigma_{\perp}\rangle$ should be along [$010$] direction based on real-space magnetic configuration. The orientations in the description of our experimental results are relative to the lattice vectors of the pseudocubic unit cell.
    
    Scanning CPGE microscopy is conducted to investigate the consistency between the equivalent spin-polarization direction and our symmetry analysis. When the sample is illuminated with circularly polarized light, the selective excitation of these spin-split states generates a directional DC photocurrent, $J_{CPGE}$, which can be phenomenologically described as $J_{CPGE}=i\beta_{ab}(\boldsymbol{E}(\omega)\times{\boldsymbol{E}^{*}(\omega)})_{b}$, %$J_{CPGE}=\sigma_{ab}E_a(\omega)E_b(\omega)$  
    where $\beta_{ab}$ is the CPGE tensor sensitive to the parity-breaking magnetic order, $\boldsymbol{E}(\omega)$ is the electrical field of illuminated light with frequency $\omega$. {As shown in Fig.~\ref{fig:2}c,} the anisotropic CPGE measurements are performed on {a device consist of two parallel $50~\mu\mathrm{m}\times1.5~\mu\mathrm{m}$ detector electrode strips, with the devices rotated in $15^\circ$ increments (from $0^\circ$ to $180^\circ$) under fixed laser excitation at the wavelength of $405~\mathrm{nm}$ and a power of $2.412~\mathrm{mW}$. The spin photocurrent are detected perpendicular to the electrode strips. $\varphi$ represents the angle between the laser polarization direction and the $\lambda/4$ plate axis, while $\theta$ indicates the angle between the direction of the measured photocurrents and the [$100$] orientation of BFO.} 
    
    Fig.~\ref{fig:2}d presents the photocurrent as a function of $\varphi$ for device orientations at $90^\circ$ and $270^\circ$. The total photocurrent contains contributions from both linearly and circularly polarized light components, where the circularly polarized contribution corresponds to the CPGE signal. For the device aligned at $90^\circ$, the photocurrent at $\varphi$ = $45^\circ$ is smaller than that at $135^\circ$, yielding a negative CPGE signal. In contrast, for the device oriented at $270^\circ$, the photocurrent at $45^\circ$ is larger than that at $135^\circ$, resulting in a positive CPGE signal. Fig.~\ref{fig:2}e displays the angular dependence of the CPGE signal and the linear polarization-induced photocurrent as a function of $\theta$. The linear polarization contribution exhibits extrema at $0^\circ$ and $180^\circ$, consistent with the in-plane projection direction of the ferroelectric polarization. The deduced CPGE current exhibits a clear sinusoidal angular dependence, with two distinct peaks and valleys, corresponding to the net spin polarization direction. The measured peak positions (at $90^\circ$ and $270^\circ$) match exactly with the direction of $p$-wave spin polarization derived from the cycloidal spin order in BFO—specifically, the net spin polarization direction is aligned along [$010$], confirming the coexistence of ferroelectricity and $p$-wave magnetism at room temperature.
    
%%    To distinguish this $p$-wave signal from conventional SOC-driven effects (such as the Rashba-Edelstein effect), we measured the CPGE spectrum from $400~\mathrm{nm}$ to $800~\mathrm{nm}$. We observed a pronounced resonance peak near $680~\mathrm{nm}$ ($\sim 1.82~\mathrm{eV}$), which aligns with the predicted transitions between the spin-split $p$-like states of the Fe $3d$-$2p$ hybridized bands. The magnitude of the CPGE dichroism reaches approximately 12 at room temperature—significantly higher than values typically observed in Rashba semiconductors, underscoring the strength of the non-relativistic exchange-driven splitting. 
    
\begin{figure}[t!]
    	\centering
    	\includegraphics[width=1\textwidth]{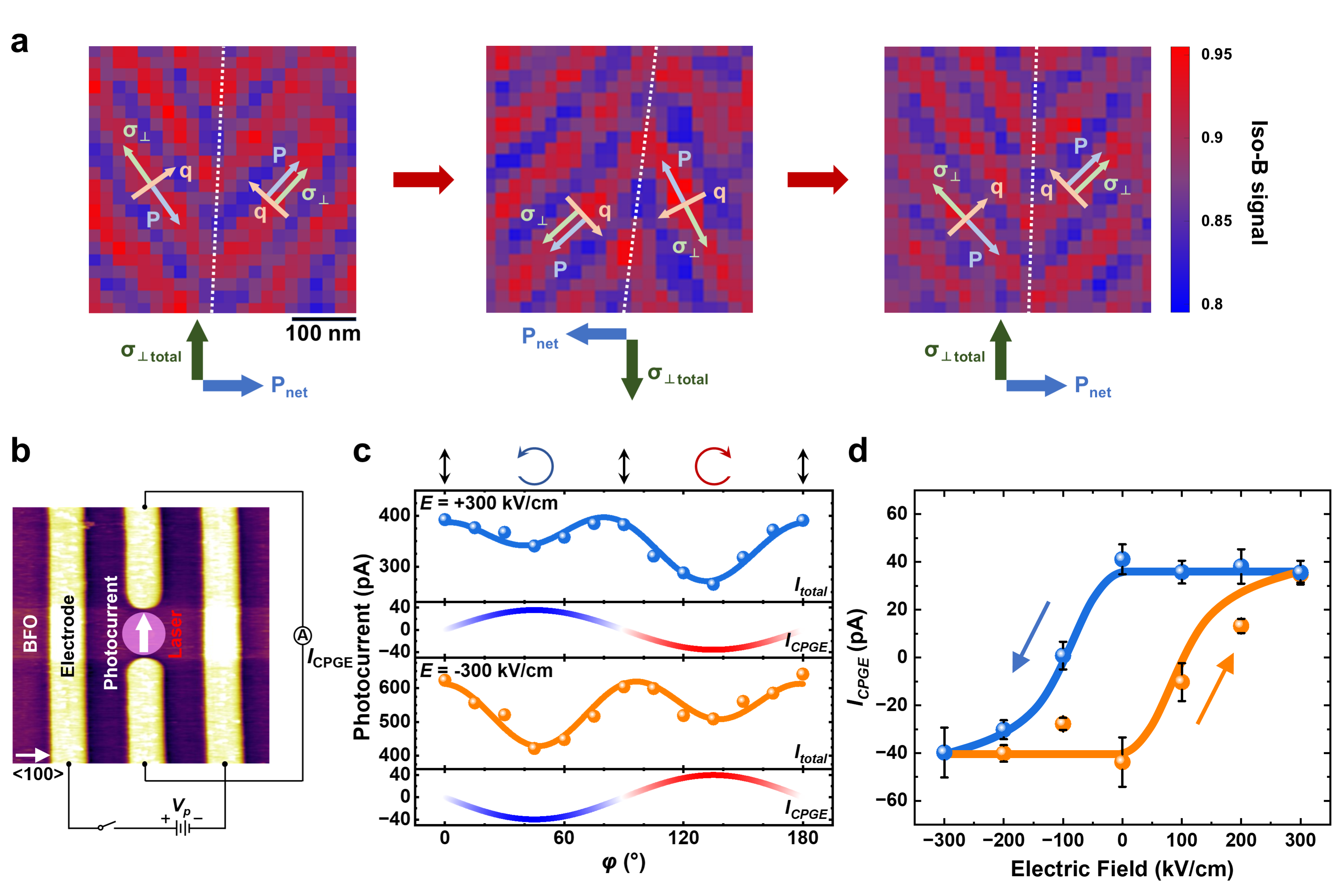}
    	\caption{\textbf{Deterministic electrical switching of $p$-wave states. a,} Electric field control of spin cycloid structure in BFO. Both ferroelectric polarization and the magnetic chirality of BFO’s cycloidal spin order are flipped under inverse saturated electric field observed by SNVM. \textbf{b, }Schematic of the experimental setup for electric-field-modulated CPGE effect measurement. A 405~nm laser with polarization defined by a $\lambda/4$ plate irradiates the BFO surface, with the definitions of $\varphi$ and $\theta$ consistent with those in Fig.~\ref{fig:2}c. The outer parallel electrode strips are employed to apply an external electric field to the BFO thin film, while the inner head-to-head electrode strips are used for photocurrent detection. The measured photocurrent direction is perpendicular to the applied electric field direction. \textbf{c, }Photocurrent line-scans under inverse electric field pulses (E = $\pm$300 kV/cm). $I_{CPGE}$ exhibit a non-volatile $180^\circ$ phase flip, which reveals that the CPGE effect contribution can be electrically reversed. \textbf{d, }Electric field dependence of $I_{CPGE}$ that displays a hysteresis loop. 
        } 
    	\label{fig:3}
    	%\vspace{80pt}
    \end{figure}

    The most compelling feature of ferroelectric $p$-wave magnetism in BFO is its susceptibility to external electric fields, which originates from the intrinsic coupling between ferroelectric polarization and magnetic chirality. In stark contrast to other magnetic systems, the magnetic chirality of BFO’s cycloidal spin order is not a free parameter; instead, it is tightly locked to the ferroelectric polarization $\mathbf{P}$ through the inverse Dzyaloshinskii-Moriya (DM) interaction, a key coupling mechanism that enables the electrical manipulation of magnetic order~\cite{bhattacharjee2014ultrafast,meisenheimer2024switching,pal2026bulk}. To demonstrate this electrical tunability, we applied an in-plane electric field pulse ($E\sim300~\mathrm{kV/cm}$) to the BFO sample, which successfully induced a full and reversible flip of the ferroelectric polarization $\mathbf{P}$. As illustrated in Fig.~\ref{fig:3}a, from the NV-center images obtained before and after electric-field switching, we find that the ferroelectric direction within each ferroelectric domain undergoes a $71^\circ$ reorientation (corresponding to a $90^\circ$ in-plane rotation), accompanied by a $90^\circ$ rotation of the $\mathbf{q}$ vector. Accordingly, cycloid’s rotational direction (handedness) is flipped from clockwise to counterclockwise (or vice versa) upon polarization reversal and the effective spin-splitting undergoes a $180^\circ$ reversal. Based on symmetry analysis, the net effective spin-splitting direction should be reversed from the [$010$] orientation to the [$0\overline{1}0$] orientation. %As illustrated in Fig.~\ref{fig:3}a, this electrical pulse triggers a concurrent and deterministic reversal of both the ferroelectric polarization and the magnetic cycloid’s handedness: our experimental reconstruction of the cycloidal structure (inset of Fig.~\ref{fig:3}a) clearly shows that the cycloid’s rotational direction (handedness) is flipped from clockwise to counterclockwise (or vice versa) upon polarization reversal. 
    
    As illustrated in Fig.~\ref{fig:3}b, to quantify the effect of cycloid handedness reversal on $p$-wave spin polarization, we perform CPGE measurements over the same sample region after applying electric field pulses of 300~$\textrm{kV/cm}$ and $-300~\textrm{kV/cm}$, with the CPGE detection direction perpendicular to the applied electric field. Figure~\ref{fig:3}c displays the photocurrent evolution as a function of the angle $\varphi$. After applying a 300~$\textrm{kV/cm}$ pulse, the photocurrent at $45^\circ$ is larger than that at $135^\circ$, corresponding to a positive CPGE signal. In contrast, following the $-300~\textrm{kV/cm}$ pulse, the photocurrent at $45^\circ$ becomes smaller than that at $135^\circ$, giving a negative CPGE signal. The results reveal a clear and deterministic sign reversal of the CPGE photocurrent: upon switching the electric field from positive to negative (accompanied by ferroelectric polarization reversal and cycloid handedness flip), the CPGE current changes sign from positive to negative (or vice versa). This sign reversal directly manifests the reversal of the  $p$-wave spin polarization direction, given that the CPGE current is linearly dependent on the spin polarization orientation.
    %To quantify the effect of cycloid handedness reversal on $p$-wave spin splitting, we measured CPGE over the same area of the sample before and after applying the electric field pulse of 350 kV/cm. The schematics is shown in Fig.~\ref{fig:3}b, as the CPGE detecting direction is perpendicular to applied electric field. The results show a clear and deterministic sign reversal of the CPGE photocurrent: when the electric field is switched from positive to negative (corresponding to polarization reversal and cycloid handedness flip), the CPGE current changes from positive to negative (or vice versa). This sign reversal directly reflects the reversal of the equivalent $p$-wave spin splitting direction, as the CPGE current is linearly dependent on the spin splitting orientation. 
    Further confirmation of this reversible tuning is provided by Fig.~\ref{fig:3}d, which shows the CPGE current as a function of the applied electric field ({\textit{i.e.}}, the CPGE-electric field loop). The loop exhibits a clear hysteretic behavior that is well-correlated with the ferroelectric polarization loop of BFO, with the CPGE current sign flipping precisely when the electric field crosses the coercive field of the ferroelectric domain switching. This hysteretic response not only confirms the reversible electric-field control of antiferromagnetic $p$-wave spin polarization but also represents the realization of a voltage-controlled odd-parity spin splitting magnet operating at room temperature.
    
    In summary, we experimentally demonstrate that BFO's cycloidal spin order is a room-temperature ferroelectric $p$-wave magnet, with non-relativistic spin splitting originated from chiral cycloid exchange interactions (not spin-orbit coupling). The $p$-wave state in BFO, driven by its chiral cycloid, allows for a direct coupling between magnetic handedness and carrier spin, which we have probed via CPGE. This effective spin splitting can be switched with electric-field at room temperature—thus realizing voltage-controlled odd-parity spin splitting in a room-temperature magnet, a key advance that opens new avenues for the development of low-power, symmetry-engineered spintronic devices.

  \newpage
    
    \FloatBarrier

    \FloatBarrier

    \FloatBarrier
    
    \clearpage
    \noindent\textbf{References}
    \bibliographystyle{naturemag}
    \bibliography{bibliography}

@article{huang2024,
  title={Manipulating chiral spin transport with ferroelectric polarization},
  author={Huang, Xiaoxi and Chen, Xianzhe and Li, Yuhang and Mangeri, John and Zhang, Hongrui and Ramesh, Maya and Taghinejad, Hossein and Meisenheimer, Peter and Caretta, Lucas and Susarla, Sandhya and others},
  journal={Nat. Mater.},
  volume={23},
  number={7},
  pages={898--904},
  year={2024},
  publisher={Nature Publishing Group UK London}
}

@article{yamada2025,
  author={Yamada, Rinsuke and Birch, Max T. and Baral, Priya R. and Okumura, Shun and Nakano, Ryota and Gao, Shang and Ezawa, Motohiko and Nomoto, Takuya and Masell, Jan and Ishihara, Yuki and Kolincio, Kamil K. and Belopolski, Ilya and Sagayama, Hajime and Nakao, Hironori and Ohishi, Kazuki and Ohhara, Takashi and Kiyanagi, Ryoji and Nakajima, Taro and Tokura, Yoshinori and Arima, Taka-hisa and Motome, Yukitoshi and Hirschmann, Moritz M. and Hirschberger, Max},
  title={A metallic $p$-wave magnet with commensurate spin helix},
  journal={Nature},
  year={2025},
  volume={646},
  number={8086},
  pages={837-842},
  publisher={Nature Publishing Group UK London}
}

@article{song2025,
  author={Song, Qian and Stavri{\'{c}}, Srdjan and Barone, Paolo and Droghetti, Andrea and Antonenko, Daniil S. and Venderbos, J{\"o}rn W. F. and Occhialini, Connor A. and Ilyas, Batyr and Erge{\c{c}}en, Emre and Gedik, Nuh and Cheong, Sang-Wook and Fernandes, Rafael M. and Picozzi, Silvia and Comin, Riccardo},
  title={Electrical switching of a $p$-wave magnet},
  journal={Nature},
  year={2025},
  volume={642},
  number={8066},
  pages={64-70},
  publisher={Nature Publishing Group UK London}
}

@article{kurz2004ab,
  title={Ab initio treatment of noncollinear magnets with the full-potential linearized augmented plane wave method},
  author={Kurz, Ph and F{\"o}rster, F and Nordstr{\"o}m, Lars and Bihlmayer, Gustav and Bl{\"u}gel, Stefan},
  journal={Physical Review B},
  volume={69},
  number={2},
  pages={024415},
  year={2004},
  publisher={APS}
}

@article{heide2009describing,
  title={Describing Dzyaloshinskii--Moriya spirals from first principles},
  author={Heide, M and Bihlmayer, G and Bl{\"u}gel, S},
  journal={Physica B: Condensed Matter},
  volume={404},
  number={18},
  pages={2678--2683},
  year={2009},
  publisher={Elsevier}
}

@article{wortmann2023fleur,
  title={Fleur},
  author={Wortmann, Daniel and Michalicek, Gregor and Baadji, Nadjib and Beida, Wejdan and Betzinger, Markus and Bihlmayer, Gustav and Bornhake, Thomas and Br{\"o}der, Jens and Burnus, Tobias and Enkovaara, Jussi and others},
  journal={Zenodo},
  year={2023}
}

@article{priessnitz2026,
  title={Ferroelectric $ p $-wave magnets},
  author={Priessnitz, Jan and Hellenes, Anna Birk and Comin, Riccardo and {\v{S}}mejkal, Libor},
  journal={arXiv preprint arXiv:2603.19107},
  year={2026}
}

@article{jungwirth2026,
  title={Symmetry, microscopy and spectroscopy signatures of altermagnetism},
  author={Jungwirth, Tomas and Sinova, Jairo and Fernandes, Rafael M and Liu, Qihang and Watanabe, Hikaru and Murakami, Shuichi and Nakatsuji, Satoru and {\v{S}}mejkal, Libor},
  journal={Nature},
  volume={649},
  number={8098},
  pages={837--847},
  year={2026},
  publisher={Nature Publishing Group UK London}
}

@article{hellenes2024,
  title={$P$-wave magnets}, 
  author={Anna Birk Hellenes and Tomáš Jungwirth and Rodrigo Jaeschke-Ubiergo and Atasi Chakraborty and Jairo Sinova and Libor Šmejkal},
  year={2024},
  pages={2309.01607},
  journal={arXiv preprint}
}

@article{brekke2024,
  title = {Minimal Models and Transport Properties of Unconventional $p$-Wave Magnets},
  author = {Brekke, Bj\o{}rnulf and Sukhachov, Pavlo and Giil, Hans Gl\o{}ckner and Brataas, Arne and Linder, Jacob},
  journal = {Phys. Rev. Lett.},
  volume = {133},
  issue = {23},
  pages = {236703},
  numpages = {9},
  year = {2024},
  month = {Dec},
  publisher = {American Physical Society}
}

@article{ezawa2025,
  title = {Purely electrical detection of the spin-splitting vector in $p$-wave magnets based on linear and nonlinear conductivities},
  author = {Ezawa, Motohiko},
  journal = {Phys. Rev. B},
  volume = {112},
  issue = {12},
  pages = {125412},
  numpages = {10},
  year = {2025},
  month = {Sep},
  publisher = {American Physical Society}
}

@article{ezawa2024,
  title = {Topological insulators and superconductors based on $p$-wave magnets: Electrical control and detection of a domain wall},
  author = {Ezawa, Motohiko},
  journal = {Phys. Rev. B},
  volume = {110},
  issue = {16},
  pages = {165429},
  numpages = {7},
  year = {2024},
  month = {Oct},
  publisher = {American Physical Society}
}

@article{jungwirth2025,
  title={From superfluid $^{3}${H}e to altermagnets}, 
  author={T. Jungwirth and R. M. Fernandes and E. Fradkin and A. H. MacDonald and J. Sinova and L. Smejkal},
  year={2025},
  pages={2411.00717},
  journal={arXiv preprint}
}

@article{bhowal2024,
  title = {{Ferroically Ordered Magnetic Octupoles in $d$-Wave Altermagnets}},
  author = {Bhowal, Sayantika and Spaldin, Nicola A.},
  journal = {Phys. Rev. X},
  volume = {14},
  issue = {1},
  pages = {011019},
  numpages = {19},
  year = {2024},
  month = {Feb},
  publisher = {American Physical Society}
}

@article{zhang2024,
  title={Predictable gate-field control of spin in altermagnets with spin-layer coupling},
  author={Zhang, Run-Wu and Cui, Chaoxi and Li, Runze and Duan, Jingyi and Li, Lei and Yu, Zhi-Ming and Yao, Yugui},
  journal={Phys. Rev. Lett.},
  volume={133},
  number={5},
  pages={056401},
  year={2024},
  publisher={APS}
}

@article{gu2025,
  title={Ferroelectric switchable altermagnetism},
  author={Gu, Mingqiang and Liu, Yuntian and Zhu, Haiyuan and Yananose, Kunihiro and Chen, Xiaobing and Hu, Yongkang and Stroppa, Alessandro and Liu, Qihang},
  journal={Phys. Rev. Lett.},
  volume={134},
  number={10},
  pages={106802},
  year={2025},
  publisher={APS}
}

@article{duan2025,
  title={Antiferroelectric altermagnets: Antiferroelectricity alters magnets},
  author={Duan, Xunkai and Zhang, Jiayong and Zhu, Ziye and Liu, Yuntian and Zhang, Zhenyu and {\v{Z}}uti{\'c}, Igor and Zhou, Tong},
  journal={Phys. Rev. Lett.},
  volume={134},
  number={10},
  pages={106801},
  year={2025},
  publisher={APS}
}

@article{wang2026,
  title={Electric field switching of altermagnetic spin-splitting in multiferroic skyrmions}, 
  author={Gui Wang and Yuhang Li and Bin Li and Xianzhe Chen and Jianting Dong and Weizhao Chen and Xiaobing Chen and Naifu Zheng and Maosen Guo and Aomei Tong and Hua Bai and Hongrui Zhang and Yifan Gao and Kaiwen Shen and Jiangyuan Zhu and Jiahao Han and Yingfen Wei and Hao Jiang and Xumeng Zhang and Ming Wang and Kebiao Xu and Wu Shi and Pengfei Wang and Jia Zhang and Qihang Liu and Cheng Song and Qi Liu and Xincheng Xie and Ming Liu},
  year={2026},
  pages={2601.06738},
  journal={arXiv preprint} 
}

@article{hodt2025,
  title = {Fate of $p$-wave spin polarization in helimagnets with Rashba spin-orbit coupling},
  author = {Hodt, Erik W. and Bentmann, Hendrik and Linder, Jacob},
  journal = {Phys. Rev. B},
  volume = {111},
  issue = {20},
  pages = {205416},
  numpages = {8},
  year = {2025},
  month = {May},
  publisher = {American Physical Society}
}

@article{ahn2019,
  title = {Antiferromagnetism in {RuO}$_2$ as $d$-wave {P}omeranchuk instability},
  author = {Ahn, Kyo-Hoon and Hariki, Atsushi and Lee, Kwan-Woo and Kune\ifmmode \check{s}\else \v{s}\fi{}, Jan},
  journal = {Phys. Rev. B},
  volume = {99},
  issue = {18},
  pages = {184432},
  numpages = {5},
  year = {2019},
  month = {May},
  publisher = {American Physical Society}
}

@article{maeda2024,
  author = {Maeda ,Kazuki and Lu ,Bo and Yada ,Keiji and Tanaka ,Yukio},
  title = {Theory of Tunneling Spectroscopy in Unconventional p-Wave Magnet-Superconductor Hybrid Structures},
  journal = {J. Phys. Soc. Jpn.},
  volume = {93},
  number = {11},
  pages = {114703},
  year = {2024}
}

@article{bhattacharjee2014ultrafast,
  title={Ultrafast switching of the electric polarization and magnetic chirality in BiFeO 3 by an electric field},
  author={Bhattacharjee, Satadeep and Rahmedov, Dovran and Wang, Dawei and {\'I}niguez, Jorge and Bellaiche, Laurent},
  journal={Physical Review Letters},
  volume={112},
  number={14},
  pages={147601},
  year={2014},
  publisher={APS}
}

@article{pal2026bulk,
  title={Bulk-Like Spin Cycloid and Fast Switching in Freestanding BiFeO3},
  author={Pal, Pratap and Deng, Xiaoxi and Kim, Gi-Yeop and Lee, Kyoungjun and Lin, Jheng-Cyuan and Mandal, Prithwijit and Yang, Jiangfeng and Steadman, Paul and Manuel, Pascal and Johnson, Roger D and others},
  journal={Advanced Functional Materials},
  pages={e77954},
  year={2026},
  publisher={Wiley Online Library}
}

@article{meisenheimer2024switching,
  title={Switching the spin cycloid in BiFeO3 with an electric field},
  author={Meisenheimer, Peter and Moore, Guy and Zhou, Shiyu and Zhang, Hongrui and Huang, Xiaoxi and Husain, Sajid and Chen, Xianzhe and Martin, Lane W and Persson, Kristin A and Griffin, Sin{\'e}ad and others},
  journal={Nature Communications},
  volume={15},
  number={1},
  pages={2903},
  year={2024},
  publisher={Nature Publishing Group UK London}
}

@article{mitscherling2026microscopic,
  title={Microscopic origin of $ p $-wave magnetism},
  author={Mitscherling, Johannes and Priessnitz, Jan and Geschner, Clara K and {\v{S}}mejkal, Libor},
  journal={arXiv preprint arXiv:2603.09736},
  year={2026}
}

@article{jungwirth2025altermagnetism,
  title={Altermagnetism: An unconventional spin-ordered phase of matter},
  author={Jungwirth, Tom{\'a}{\v{s}} and Fernandes, Rafael M and Fradkin, Eduardo and MacDonald, Allan H and Sinova, Jairo and {\v{S}}mejkal, Libor},
  journal={Newton},
  volume={1},
  number={6},
  year={2025},
  publisher={Elsevier}
}

@article{elcoro2026spin,
  title={Spin point group symmetry and classification of non-relativistic spin splitting in non-collinear magnetic structures: Identification of high-order spin splitting types (l= 5, 7, and 9)},
  author={Elcoro, Luis and Etxebarria, Jesus and Perez-Mato, J Manuel and Tasci, Emre S},
  journal={arXiv preprint arXiv:2606.19254},
  year={2026}
}

@article{feng2022,
  title={{An anomalous Hall effect in altermagnetic ruthenium dioxide}},
  author={Feng, Zexin and Zhou, Xiaorong and {\v{S}}mejkal, Libor and Wu, Lei and Zhu, Zengwei and Guo, Huixin and Gonz{\'a}lez-Hern{\'a}ndez, Rafael and Wang, Xiaoning and Yan, Han and Qin, Peixin and others},
  journal={Nat. Electron.},
  volume={5},
  number={11},
  pages={735--743},
  year={2022},
  publisher={Nature Publishing Group UK London}
}

@article{krempasky2024,
  title={{Altermagnetic lifting of Kramers spin degeneracy}},
  author={Krempask{\`y}, Juraj and {\v{S}}mejkal, L and D’souza, SW and Hajlaoui, M and Springholz, G and Uhl{\'\i}{\v{r}}ov{\'a}, K and Alarab, F and Constantinou, PC and Strocov, V and Usanov, D and others},
  journal={Nature},
  volume={626},
  number={7999},
  pages={517--522},
  year={2024},
  publisher={Nature Publishing Group UK London}
}

@article{smejkal2023,
  title={{Chiral magnons in altermagnetic RuO$_2$}},
  author={{\v{S}}mejkal, Libor and Marmodoro, Alberto and Ahn, Kyo-Hoon and Gonz{\'a}lez-Hern{\'a}ndez, Rafael and Turek, Ilja and Mankovsky, Sergiy and Ebert, Hubert and D’Souza, Sunil W and {\v{S}}ipr, Ond{\v{r}}ej and Sinova, Jairo and others},
  journal={Phys. Rev. Lett.},
  volume={131},
  number={25},
  pages={256703},
  year={2023},
  publisher={APS}
}

@article{amin2024,
  author={Amin, O. J. and Dal Din, A. and Golias, E. and Niu, Y. and Zakharov, A. and Fromage, S. C. and Fields, C. J. B. and Heywood, S. L. and Cousins, R. B. and Maccherozzi, F. and Krempask{\'y}, J. and Dil, J. H. and Kriegner, D. and Kiraly, B. and Campion, R. P. and Rushforth, A. W. and Edmonds, K. W. and Dhesi, S. S. and {\v{S}}mejkal, L. and Jungwirth, T. and Wadley, P.},
  title={Nanoscale imaging and control of altermagnetism in MnTe},
  journal={Nature},
  year={2024},
  month={Dec},
  day={01},
  volume={636},
  number={8042},
  pages={348-353}
}

@article{reichlova2024,
  title={{Observation of a spontaneous anomalous Hall response in the Mn$_5$Si$_3$ d-wave altermagnet candidate}},
  author={Reichlova, Helena and Lopes Seeger, Rafael and Gonz{\'a}lez-Hern{\'a}ndez, Rafael and Kounta, Ismaila and Schlitz, Richard and Kriegner, Dominik and Ritzinger, Philipp and Lammel, Michaela and Leivisk{\"a}, Miina and Birk Hellenes, Anna and others},
  journal={Nat. Commun.},
  volume={15},
  number={1},
  pages={4961},
  year={2024},
  publisher={Nature Publishing Group UK London}
}

@article{chen2025,
  title={Electrical switching of altermagnetism},
  author={Chen, Yiyuan and Liu, Xiaoxiong and Lu, Hai-Zhou and Xie, XC},
  journal={Phys. Rev. Lett.},
  volume={135},
  number={1},
  pages={016701},
  year={2025},
  publisher={APS}
}

@article{ederer2005,
  title={{Weak ferromagnetism and magnetoelectric coupling in bismuth ferrite}},
  author={Ederer, Claude and Spaldin, Nicola A.},
  journal={Phys. Rev. B},
  volume={71},
  number={6},
  pages={060401(R)},
  year={2005},
  publisher={APS}
}

@article{rahmedov2012,
  title={{Magnetic Cycloid of {B}i{F}e{O}$_3$ from Atomistic Simulations}},
  author={Rahmedov, Dovran and Wang, Dawei and Íñiguez, Jorge and Bellaiche, Laurent},
  journal={Phys. Rev. Lett.},
  volume={109},
  pages={037207},
  year={2012},
  publisher={American Physical Society}
}

@article{chai2024,
  title={Voltage control of multiferroic magnon torque for reconfigurable logic-in-memory},
  author={Chai, Yahong and Liang, Yuhan and Xiao, Cancheng and Wang, Yue and Li, Bo and Jiang, Dingsong and Pal, Pratap and Tang, Yongjian and Chen, Hetian and Zhang, Yuejie and others},
  journal={Nat. Commun.},
  volume={15},
  number={1},
  pages={5975},
  year={2024},
  publisher={Nature Publishing Group UK London}
}

@article{albrecht2010,
  title = {{Ferromagnetism in multiferroic ${\text{BiFeO}}_{3}$ films: A first-principles-based study}},
  author = {Albrecht, D. and Lisenkov, S. and Ren, Wei and Rahmedov, D. and Kornev, Igor A. and Bellaiche, L.},
  journal = {Phys. Rev. B},
  volume = {81},
  issue = {14},
  pages = {140401},
  numpages = {4},
  year = {2010},
  month = {Apr},
  publisher = {American Physical Society}
}

@article{husain2026,
  title={{Anisotropic magnon transport in an antiferromagnetic trilayer heterostructure: is BiFeO$_3$ an altermagnet?}},
  author={Husain, Sajid and Ramesh, Maya and Song, Qian and Prokhorenko, Sergei and Ojha, Shashank Kumar and Panda, Surya Narayan and Li, Xinyan and Nahas, Yousra and Kumar, Yogesh and Gupta, Pushpendra and others},
  pages={2601.04578},
  journal={arXiv preprint},
  year={2026}
}

@article{li2023,
  title = {{Multiferroic skyrmions in ${\mathrm{BiFeO}}_{3}$}},
  author = {Li, Z. and Chirac, T. and Tranchida, J. and Garcia, V. and Fusil, S. and Jacques, V. and Chauleau, J.-Y. and Viret, M.},
  journal = {Phys. Rev. Res.},
  volume = {5},
  issue = {4},
  pages = {043109},
  numpages = {6},
  year = {2023},
  month = {Nov}
}

@article{manipatruni2019,
  author={Manipatruni, Sasikanth and Nikonov, Dmitri E. and Lin, Chia-Ching and Gosavi, Tanay A. and Liu, Huichu and Prasad, Bhagwati and Huang, Yen-Lin and Bonturim, Everton and Ramesh, Ramamoorthy and Young, Ian A.},
  title={Scalable energy-efficient magnetoelectric spin--orbit logic},
  journal={Nature},
  year={2019},
  month={Jan},
  day={01},
  volume={565},
  number={7737},
  pages={35-42}
}

@article{meisenheimer2024,
  title={Designed Spin-Texture-Lattice to Control Anisotropic Magnon Transport in Antiferromagnets},
  author={Meisenheimer, Peter and Ramesh, Maya and Husain, Sajid and Harris, Isaac and Park, Hyeon Woo and Zhou, Shiyu and Taghinejad, Hossein and Zhang, Hongrui and Martin, Lane W and Analytis, James and others},
  journal={Adv. Mater.},
  volume={36},
  number={36},
  pages={2404639},
  year={2024},
  publisher={Wiley Online Library}
}

@article{caretta2023,
  author={Caretta, Lucas and Shao, Yu-Tsun and Yu, Jia and Mei, Antonio B. and Grosso, Bastien F. and Dai, Cheng and Behera, Piush and Lee, Daehun and McCarter, Margaret and Parsonnet, Eric and K. P, Harikrishnan and Xue, Fei and Guo, Xiangwei and Barnard, Edward S. and Ganschow, Steffen and Hong, Zijian and Raja, Archana and Martin, Lane W. and Chen, Long-Qing and Fiebig, Manfred and Lai, Keji and Spaldin, Nicola A. and Muller, David A. and Schlom, Darrell G. and Ramesh, Ramamoorthy},
  title={Non-volatile electric-field control of inversion symmetry},
  journal={Nat. Mater.},
  year={2023},
  month={Feb},
  day={01},
  volume={22},
  number={2},
  pages={207-215}
}
    
    \noindent\textbf{Methods}
\\
    \noindent\textbf{DFT calculations}\\
   Our density functional theory calculations were performed in FLEUR~\upcite{wortmann2023fleur} within the Full-Potential Linearized Augmented-Plane-Wave (FPLAPW) method.  The muffin-tin radii were set to $2.31~a_0$, $2.8~a_0$, and $1.3~a_0$ for Fe, Bi, and O, respectively, where $a_0$ is the Bohr radius. The plane-wave and density cutoffs were chosen as $k_{\mathrm{max}}=4.5~a_0^{-1}$ and  $G_{\mathrm{MAX}}=10^{-5} \mathrm{m}e/a_0^3$, respectively. The self consistent calculations were performed on a $5\times5\times5$ $k$-mesh with a convergence criteria of $10^{-7}$~Hartree for the total energy and $10^{-5} \mathrm{m}e/a_0^3$ for the charge density. The non-self-consistent calculations for the spin texture were performed on a $75\times75\times1$ 2D grid within the generalized Bloch theorem for nonrelativistic spin spirals~\upcite{kurz2004ab}. The relativistic band structure was calculated perturbatively with the projection of the spin-orbit coupling Hamiltonian onto the generalized Bloch eigenstates, as implemented in FLEUR with the magnetic force theorem~\upcite{heide2009describing}.
   
    %Theoretical calculations were carried out using first‑principles methods implemented in the Vienna ab initio simulation package (VASP)\upcite{kresse1993}. The projector augmented wave (PAW) pseudo potential\upcite{blochl1994} was employed together with the Perdew–Burke–Ernzerhof (PBE) generalized gradient approximation (GGA) for the exchange–correlation functional\upcite{perdew1996}. A $2\times2\times1$ supercell was constructed with lattice constants a=b=$15.96~\textrm{\AA}$ and c=$7.98~\textrm{\AA}$. To ensure convergence of the self‑consistent calculations, an energy cutoff of $600$~eV and an energy convergence criterion of $10^{-5}$~eV were adopted. A $k$‑point mesh of $3\times3\times5$ was used to sample the Brillouin zone uniformly, corresponding to the given lattice parameters.\\

    \noindent\textbf{Sample and device fabrication}\\
    The BFO film used in this study is grown by pulsed laser deposition on DyScO${}_3$ substrate. A 100~nm BFO layer is grown at a substrate temperature 800~$^\circ$C and oxygen partial pressure 0.015~mbar. After the growth, the sample is cooled to room temperature in an oxygen environment of 300~mbar at a cooling rate of 20~$^\circ$C/min. The devices are fabricated using photolithography. Afterwards, the Pt layer of 15~nm is deposited by physical vapor deposition.\\
\\
    \noindent\textbf{Scanning NV magnetometry measurements}\\
    Scanning nitrogen-vacancy magnetometry (SNVM) was carried out under ambient conditions using a commercial all-diamond scanning probe tip hosting individual NV defects from CIQTEK, which was fitted to a tuning-fork based atomic force microscope (AFM) integrated with a confocal setup tailored for single-NV defect spectroscopy.
    %Scanning-NV magnetometry was performed under ambient conditions with commercial all-diamond scanning probe tips containing single NV defects (SNVM, CIQTEK). The tip was integrated into a tuning-fork based atomic force microscope (AFM) combined with a confocal microscope optimized for single NV defect spectroscopy. 
    Magnetic fields emanating from the sample are detected by recording the Zeeman shift of the NV defect's electronic spin sublevels through optical detection of the electron spin resonance. Experiments were performed with a NV-to-sample distance of d = $47.7\pm1.3$~nm and a bias magnetic field of 2~mT applied along the NV quantization axis which is at an angle of $\theta_{NV}=59.8\pm1.9^\circ$ and $\psi_{NV}=180\pm 1^\circ$ with respect to the surface normal.
    %Scanning nitrogen-vacancy magnetometry (SNVM) was carried out under ambient conditions using a commercial all-diamond scanning probe tip hosting individual NV defects from CIQTEK, which was fitted to a tuning-fork based atomic force microscope (AFM) integrated with a confocal setup tailored for single-NV defect spectroscopy. Local magnetic stray fields from the sample were characterized by monitoring the Zeeman splitting of the NV electronic spin states through optically detected magnetic resonance. All measurements were acquired at a tip–sample separation of d = $47.7\pm1.3$~nm, while a 2 mT static bias field was aligned along the NV quantization axis. This axis adopts angular orientations of $\theta_{NV}=59.8\pm1.9^\circ$ and $\psi_{NV}=180\pm 1^\circ$ referenced to the sample surface normal.
    \\
\\
    \noindent\textbf{Photocurrent measurements}\\
    Photocurrent measurement was conducted using a home-built confocal microscope. A 405~nm solid-state laser (CNI Optoelectronics Technology Co., Ltd.) was used as the excitation source. The laser power was attenuated to 2.412~mW using a density filter and the laser spot %(diameter \textless 2 $\mu m$) 
    was focused on the sample at normal incidence using a $\times 50$ Olympus objective (0.5~NA). The polarization state of the 405~nm laser was controlled by rotating a quarter-wave plate to switch between left-handed ($\varphi=45^\circ$) and right-handed ($\varphi=135^\circ$) circular polarization. Photocurrent mapping was carried out by scanning the laser spot across the sample using a two-axis galvanometer. A chopper (Sine Scientific Instruments Co., Ltd.) modulated the light at 133~Hz to facilitate lock-in detection. The photocurrent signal was amplified by an SR570 current preamplifier (Stanford Research Systems) and collected using an SR830 lock-in amplifier (Stanford Research Systems). The ferroelectric polarity of the BFO sample was switched by applying a poling voltage through a Keithley 2400 source meter (Tektronix, Inc.). All photocurrent measurements presented in the main text were performed at room temperature under ambient conditions.\\
\\
    %\noindent\textbf{Ferroelectric polarization switching}\\
    %The ferroelectric polarizations were switched using a ferroelectric tester (Precision Multiferroic, Radiant Technologies), with a fixed frequency of $10 kHz$ under room temperature. Two probes were placed on the Pt layer. \\
\\
    %{\noindent\textbf{AAA}
%\\
%\noindent AAA.

    \noindent\textbf{Data availability}\\
    The data that support the findings of this study are available from the corresponding author on reasonable request.\\
    
    \noindent\textbf{Acknowledgements}\\
    We are grateful for the fruitful discussions with Prof. Qihang Liu, Dr. Xiaobin Chen, Prof. Junwei Liu and Prof. Ramamoorthy Ramesh. N.H., W.H.C., J.P. and L.S. acknowledge support from ERC Starting Grant No. 101095925.   \\    
    
    \noindent\textbf{Author contributions}\\
    X.C. supervised this study. K.S. and G.W. carried out the synthesis of heterostructures and fabricated the devices. G.W. carried out the NV-imaging measurements. K.S. and B.L. carried out the photocurrent measurements. N.H., W.H.C., J.P., J.D., and L.S. performed the theoretical calculations and analysis. Y.G., Y.W., H.J., X.Z., M.W., J.Z., W.S., Q.L. and M.L. gave suggestions on the experiments. K.S., W.H.C., L.S., and X.C. co-wrote the manuscript. All authors discussed the results and contributed to the revision of the manuscript.\\
  
    \noindent\textbf{Competing interests}\\
    The authors declare no competing interests.\\

    \noindent\textbf{Additional information}\\ Correspondence and requests for materials should be addressed to X.C.. 
    Reprints and permissions information is available at http://www.nature.com/reprints.

    %\renewcommand\thefigure{Extended Figure \arabic{figure}}
    %\setcounter{figure}{0}
    %\captionsetup[figure]{labelfont={bf}, name={Extended Data Fig.}, labelsep=period}

    %\newpage
    %\begin{figure}[t!]
    %	\centering
    %	\includegraphics[width=1\textwidth]{Extended_Data_Figure1.png}
    %    \caption{\textbf{Difference between ferromagnetic and antiferromagnetic cycloidal spin structure. a-b, }Schematic of the ferromagnetic cycloidal spin structure with $L = 2$ in \textbf{a} and $L = 3$ in\textbf{b}. The ferromagnetic cycloid possesses only a single magnetic sublattice. Consequently, it preserves $\mathcal{T}\tau$ symmetry only for even values of $L$. \textbf{c-d, }Same as \textbf{a,b} but for the antiferromagnetic cycloidal spin structure. In contrast, the antiferromagnetic cycloid comprises two magnetic sublattices, which lifts the constraint, allowing the $\mathcal{T}\tau$ symmetry to exist for both odd and even $L$.
    %    }
    %	\label{fig:Eig1}
    %\end{figure}

    %\begin{figure}[t!]
    	%\centering
    	%\includegraphics[width=1\textwidth]{Extended_Data_Figure2.pdf}
        %\caption{\textbf{Another hysteresis loop of electric field dependence of $I_{CPGE}$ measured in a BFO/Pt bilayer device.} 
       % }
    	%\label{fig:Eig1}
    %\end{figure}
    
\end{document}